\documentclass[american]{new_tlp}

\usepackage[nodate]{datetime}
\usepackage{babel}
\usepackage{amsmath}
\usepackage{mathtools}[2011/02/12]
\usepackage{amsfonts,amssymb,stmaryrd}
\usepackage{MnSymbol}
\usepackage{mdwlist}
\usepackage[applemac]{inputenc}
\usepackage{multicol}
\usepackage{galois}

\usepackage{tikz}
\usetikzlibrary{arrows}

\usepackage[linkcolor=blue, urlcolor=blue, citecolor=blue, backref=false, final]{hyperref}
\usepackage[fancyproofs,noextended,squareitemtag]{theorems}[2013/01/24]
\makeatletter
\DeclareRobustCommand{\setlabel}[1]{\def\@currentlabel{#1}}
\makeatother

\title{Abstract Diagnosis for \tccp\ using a Linear\\ Temporal Logic}

\author[M. Comini, L. Titolo and A. Villanueva]
{MARCO COMINI, LAURA TITOLO\\
DIMI, Universit\`a degli Studi di Udine, Italy\\
\email{marco.comini@uniud.it}\\
\and ALICIA VILLANUEVA 
\\
DSIC, Universitat Polit\`ecnica de Val\`encia, Spain\\
\email{villanue@dsic.upv.es}}

\providecommand*{\monobioperator}[3]{
  \newcommand*{#1}[2]{{\ifempty{##2}{#3##1}{##1#2##2}}}
}
\makeatletter
\providecommand{\ifempty}[3]{\def\@@@temp{#1}\ifx\@@@temp\@empty#2\else#3\fi}
\makeatother

\DeclarePairedDelimiterX{\mset}[2]{\langle}{\rangle}{#1\ifempty{#2}{}{\,\delimsize\vert\, #2}}
\DeclarePairedDelimiterX{\set}[2]{\{}{\}}{#1\ifempty{#2}{}{\,\delimsize\vert\, #2}}
\monobioperator{\Lconj}{\mathrel{\dot\wedge}}{\mathrel{\dot\bigwedge}}
\monobioperator{\Ldisj}{\mathrel{\dot\vee}}{\mathrel{\dot\bigvee}}
\monobioperator{\glbCC}{\sqcap}{\bigsqcap}
\monobioperator{\glbC}{\glbCsym}{\glbCbigsym}
\monobioperator{\lubCC}{\sqcup}{\bigsqcup}
\monobioperator{\lubC}{\lubCsym}{\lubCbigsym}
\newcommand*{\Aa}{\eval[]{}}
\newcommand*{\Agents}[1][\CSys]{\mathbb{A}^{\Psyms}_{#1}}
\newcommand*{\Arule}{\parensmathoper{\mathit{A}}}
\newcommand*{\Brulel}{\parensmathoper{\mathit{B}_1}}
\newcommand*{\Bruler}{\parensmathoper{\mathit{B}_2}}

\newcommand*{\CSdom}{\mathcal{C}}
\newcommand*{\CSfalse}{\mathit{false}}
\newcommand*{\CShid}[2]{\mathop{\exists}\nolimits_{#1} #2}
\newcommand*{\CSimp}{\mathrel{\vdash}}
\newcommand*{\CSmerge}{\mathbin{\otimes}}
\newcommand*{\CSnimp}{\mathrel{\nvdash}}
\newcommand*{\CSord}{\preceq}
\newcommand*{\CStrue}{\mathit{true}}
\newcommand*{\CSys}{\mathbf{C}}
\newcommand*{\Conf}{\mathit{Conf}}
\newcommand*{\C}[1][]{\mathbb{M}_{#1}}
\newcommand*{\Dd}{\Tp[]}
\newcommand*{\Dmst}{D_{m}}
\newcommand*{\Drc}{\{\Dmst\}}
\newcommand*{\D}{\prog}
\newcommand*{\FAa}[3][]{\mathop{\mathLTL{A}_{\mathit{#1}}}\ifempty{#2}{}{\BBrackets[{#3}]{#2}}}
\newcommand*{\FDd}[1][]{\syntaxoper{\mathLTL{D}_{\mathit{#1}}}}
\newcommand*{\FI}{\mathLTL{I}}
\newcommand*{\FSym}{\mathC{F}}
\newcommand*{\Fdom}{\mathbb{F}}

\newcommand*{\FtoM}{\parensmathoper{\gamma^{\Fdom}}}
\newcommand*{\F}[2][\alpha]{\syntaxoper{\FSym^{#1}}{#2}{}}
\newcommand*{\Isym}[1]{\mathC{I}^{#1}}
\newcommand*{\I}[1][\alpha]{\Isym{#1}}
\newcommand*{\Lalways}[1]{\ifempty{#1}{\mathop{\Box}}{\mathop{\Box} #1}}
\newcommand*{\Leventually}[1]{\ifempty{#1}{\mathop{\Diamond}}{\mathop{\Diamond} #1}}
\newcommand*{\Lfalse}{\dot{\mathit{false}}}
\newcommand*{\Lhid}[2]{\mathop{\dot{\exists}}\nolimits_{#1} \ifempty{#2}{}{#2}}
\newcommand*{\Liff}[2]{\ifempty{#1}{\dot\leftrightarrow}{#1 \mathrel{\dot\leftrightarrow} #2}}
\newcommand*{\Limpl}[2]{#1 \mathrel{\dot\rightarrow} #2}
\newcommand*{\Lneg}[1]{\ifempty{#1}{\mathop{\dot{\neg}}}{\mathop{\dot{\neg}} #1}}
\newcommand*{\Lnextnum}[2]{\ifempty{#2}{\mathop{\bigcirc^{#1}}}{\mathop{\bigcirc^{#1}} #2}}
\newcommand*{\Lnext}[1]{\ifempty{#1}{\mathop{\bigcirc}}{\mathop{\bigcirc} #1}}

\newcommand*{\Ltrue}{\dot{\mathit{true}}}
\newcommand*{\Luntil}[2]{\ifempty{#1}{\mathcal{U}}{#1 \mathrel{\mathcal{U}} #2}}
\newcommand*{\MGCsym}{\mathbb{PC}}
\newcommand*{\MGC}[1][]{\MGCsym_{#1}}
\newcommand*{\ProgSym}{D}
\newcommand*{\ProgsSym}{\mathbb{D}}
\newcommand*{\Progs}[1][\CSys]{\ProgsSym^{\Psyms}_{#1}}
\newcommand*{\Psyms}{\Pi}
\newcommand*{\Qq}[2]{\Q{#2}{#1}}
\newcommand*{\Q}[2]{\ifempty{#2}{#1}{#2 \mathbin{.} #1}}
\newcommand*{\SF}{\mathLTL{S}}
\newcommand*{\Ssym}[1]{\mathC{S}^{#1}}
\newcommand*{\Sz}[1][\alpha]{\Ssym{#1}}
\newcommand*{\Tbranches}{\mathit{B}}
\newcommand*{\Tinit}[1][\Phi]{n_{#1}}
\newcommand*{\Tlabel}{\parensmathoper{\mathit{L}}}
\newcommand*{\Tname}[1][\Phi]{\mathcal{T}_{#1}}
\newcommand*{\Tnodes}{\mathit{Nodes}}
\newcommand*{\TpSym}{\mathC{D}}
\newcommand*{\Tp}[3][\alpha]{\syntaxoper{\TpSym^{#1}}{#2}{{#3}}}
\newcommand*{\Tsucc}{\mathrel{\mathit{R}}}
\newcommand*{\Ttabl}[1][\Phi]{(\Tnodes, \Tinit[#1], \Tlabel{} , \Tbranches, \Tsucc{}{} )}
\newcommand*{\Var}{\mathit{Var}}
\newcommand*{\aask}[1]{\mathop{\mathsf{ask}}\ifempty{#1}{}{(#1)\ra}}
\newcommand*{\ahiding}[3][]{\mathop{\exists^{#1}{#2}} {#3}}

\newcommand*{\anow}[3]{\tagnow\ifempty{#1}{} {\: #1 \tagthen #2\ifempty{#3}{}{\tagelse #3}}}
\newcommand*{\aparallel}[2]{#1 \parallel #2}
\newcommand*{\apcall}[2]{\mathit{#1}\ifempty{#2}{}{(\mathit{#2})}}
\newcommand*{\asat}[2]{#1 \models #2}
\newcommand*{\askip}{\mathord{\mathsf{skip}}}
\newcommand*{\asumask}[4][i]{\sum_{#1=1}^{#2}\aask{#3_{#1}}{#4}_{#1}}
\newcommand*{\atell}{\parensmathoper{\mathsf{tell}}}

\newcommand*{\botC}{\{\ecrs\}}
\newcommand*{\botF}{\Lfalse}
\newcommand*{\call}[3]{\termres{\mathit{#1}\ifempty{#2}{}{\mathit{(#2)}}}{#3}}
\newcommand*{\ccp}{\textit{ccp}}
\newcommand*{\clauseif}{\ensuremath{\mathrel{\mathord{:}\mathord{-}}}}
\newcommand*{\closed}[1][A]{\ifempty{#1}{S}{s}elf-sufficient}

\newcommand*{\cntx}[1]{{#1}^{*}}
\newcommand*{\cond}[2]{(#1 ,\allowbreak #2)}
\newcommand*{\controller}{master}
\newcommand*{\cpo}[2]{{(#1, \, \mathord{#2})}}
\newcommand*{\csC}[3]{\cs{\cond{#1}{#2}}{#3}}
\newcommand*{\csltl}[1][]{\ensuremath{\textsf{csLTL}_{#1}}}
\newcommand*{\cs}[2]{ #1 \rightarrowtail #2}
\newcommand*{\deffun}[1][A]{process}
\newcommand*{\dfn}{\coloneq}
\newcommand*{\ecrs}{\epsilon}
\newcommand*{\ed}{\boxtimes}
\newcommand*{\evalSym}{\mathC{A}}
\newcommand*{\eval}[4][\alpha]{\syntaxoperEXT{#3}{\evalSym^{#1}}{\termres{#3}{#2}}{#4}}
\newcommand*{\genrule}[2]{#1 \clauseif #2}

\newcommand*{\glbCbigsym}{\bigsqcap}
\newcommand*{\glbCsym}{\sqcap}
\newcommand*{\glbF}{\Lconj}
\newcommand*{\interpA}[1][\A]{\mathbb{I}_{#1}}
\newcommand*{\interpF}{\interpA[\Fdom]}
\newcommand*{\latticeC}{\lattice{\C}{\leqC}{\lubC{}{}}{\glbC{}{}}{\topC}{\botC}}
\newcommand*{\latticeF}{\lattice{\Fdom}{\leqF}{\lubF{}{}}{\glbF{}{}}{\topF}{\botF}}
\newcommand*{\lattice}[6]{{(#1, \, \mathord{#2}, \, \mathord{#3}, \, \mathord{#4}, \, \mathord{#5}, \, \mathord{#6})}}
\newcommand*{\leqCC}{\sqsubseteq}
\newcommand*{\leqC}{\leqCC}
\newcommand*{\leqF}{\Limpl{}{}}
\newcommand*{\lfpof}[2][]{\mathoper{lfp}_{#1}\Braces{#2}}
\newcommand*{\lfp}[1][]{\lfpof[{#1}]{}}
\newcommand*{\ltl}{\textsf{LTL}}
\newcommand*{\lubCbigsym}{\bigsqcup}
\newcommand*{\lubCsym}{\sqcup}
\newcommand*{\lubF}{\Ldisj}
\newcommand*{\lub}{\ensuremath{\mathit{lub}}}
\newcommand*{\mathC}{\mathcal}
\newcommand*{\mgc}[4][n]{\call{#2}{\mbox{$\zseq[#1]{#3}$}}{#4}}
\newcommand*{\mghead}[3][n]{\mathit{#2}\ifempty{#3}{}{(\zseq[#1]{\mathit{#3}})}}
\newcommand*{\mgrule}[4][n]{\genrule{\mghead[#1]{#2}{#3}}{#4}}
\newcommand*{\nextop}{\parensmathoper{\mathsf{next}}}
\newcommand*{\nleqCC}{\nsqsubseteq}
\newcommand*{\nleqC}{\nleqCC}

\newcommand*{\notasat}[2]{#1 \nmodels #2}
\newcommand*{\notsat}[2]{#1 \mathrel{\not\models} #2}
\newcommand*{\nra}{\not\rightarrow}
\newcommand*{\ntcc}{\textit{ntcc}}
\newcommand*{\phicont}{\parensmathoper{\phi_{\mathit{M}}}}
\newcommand*{\pltl}{\textsf{PLTL}}
\newcommand*{\programs}{sets of declarations}
\newcommand*{\program}{set of declarations}
\newcommand*{\progrules}{\progrule s}
\newcommand*{\progrule}{process declaration}
\newcommand*{\prog}[1][]{\ensuremath{\ProgSym_{\mathit{#1}}^{}}}
\newcommand*{\prule}[3]{\genrule{\mathit{#1}\ifempty{#2}{}{\mathit{(#2)}}}{#3}}
\newcommand*{\query}[1][A]{\ifempty{#1}{P}{p}rogram}
\newcommand*{\ra}{\rightarrow}
\newcommand*{\seqHid}[2]{\mathop{\bar{\exists}}\nolimits_{#1} #2}
\newcommand*{\seq}[2][n]{{#2}_{1}, \dots, {#2}_{#1}}
\newcommand*{\stutt}[1]{\parensmathoper{\mathit{stutt}}{#1}}
\newcommand*{\tagelse}{\mathrel{\mathsf{else}}}
\newcommand*{\tagnow}{\mathop{\mathsf{now}}}
\newcommand*{\tagthen}{\mathrel{\mathsf{then}}}
\newcommand*{\tccp}{\textit{tccp}}
\newcommand*{\tcc}{\textit{tcc}}
\newcommand*{\termres}[2]{{#1} \ifempty{#2}{}{}}

\newcommand*{\topC}{\mathbf{M}}
\newcommand*{\topF}{\Ltrue}
\newcommand*{\zseq}[2][n]{\vec{#2}}
\newcommand{\mathLTL}[1]{\hat{\mathcal{#1}}}
\providecommand*{\BBrackets}[2][]{\ifempty{#2}{}{\llbracket#2\rrbracket_{#1}}}
\providecommand*{\Braces}[2][]{\ifempty{#2}{}{( #2 )_{#1}}}

\providecommand*{\ie}   {i.e.,} 
\providecommand*{\mathoper}[1]{\mathop{\mathit{#1}}\nolimits}
\providecommand*{\pair}[2]{{\langle #1, \, #2 \rangle}}
\providecommand*{\parensmathoper}[2]{\ensuremath{\mathoper{#1}\ifempty{#2}{}{(#2)}}}
\providecommand*{\resp} {respectively}
\providecommand*{\st}   {s.t.}
\providecommand*{\syntaxoperEXT}[4]{\mathoper{#2}\ifempty{#1}{}{\llbracket#3\rrbracket_{#4}}}
\providecommand*{\syntaxoper}[3]{\mathoper{#1}\BBrackets[#3]{#2}}
\providecommand*{\true}{\mathit{true}}
\providecommand*{\wrt}  {w.r.t.}

\newcommand*{\sequent}[4][]{\gdef\and{\quad\hfill}
    \gdef\ands{\;\hfill\ldots\hfill\;}\global\setlabel{#1}
    {#1}\frac{\:\:#2\:\:}{\;#3\;}\;\text{\small#4}}

\renewcommand*{\P}{\prog}
    
\begin{document}
\pagestyle{plain}
\maketitle
\begin{abstract}
    Automatic techniques for program verification usually suffer the
    well-known state explosion problem.  Most of the classical approaches
    are based on browsing the structure of some form of model (which
    represents the behavior of the program) to check if a given
    specification is valid.  This implies that a part of the model has to
    be built, and sometimes the needed fragment is quite huge.
    
    In this work, we provide an alternative automatic decision method to
    check whether a given property, specified in a linear temporal logic,
    is \emph{valid} \wrt\ a \tccp\ program.  Our proposal (based on
    abstract interpretation techniques) does not require to build any model
    at all.
    Our results guarantee correctness but, as usual when using an abstract
    semantics, completeness is lost.
\end{abstract}


\section{Introduction}

The Concurrent Constraint Paradigm (\ccp, \cite{Saraswat89phd}) is a
simple, logic model which is different from other (concurrent) programming
paradigms mainly due to the notion of store-as-constraint that replaces the
classical store-as-valuation model.  It is based on an underlying
constraint system that handles constraints on variables and deals with
partial information.  Within this family, \cite{deBoerGM99} introduced the
\emph{Timed Concurrent Constraint Language} (\tccp\ in short) by adding to
the original \ccp\ model the notion of time and the ability to capture the
absence of information.  With these features, one can specify behaviors
typical of reactive systems such as \emph{timeouts} or \emph{preemption}
actions.

It is well-known that modeling and verifying concurrent systems by hand can
be an extremely hard task.  Thus, the development of automatic formal
methods is essential.  One of the most known techniques for formal
verification is model checking, that was originally introduced in
\cite{ClarkeE81,QueilleS82} to automatically check if a finite-state system
satisfies a given property.  It consisted in an exhaustive analysis of the
state-space of the system; thus the state-explosion problem is its main
drawback and, for this reason, many proposals in the literature try to
mitigate it.

All the proposals of model checking have in common that a \emph{part} of
the model of the (target) \query\ has to be built, and sometimes the needed
fragment is quite huge.  In this work, we propose a completely different
approach to the formal verification of temporal (\ltl) properties of
concurrent (reactive) systems specified in \tccp.  We formalize a method to
validate a specification of the expected behavior of a \tccp\ program $P$,
expressed by a linear temporal formula $\phi$, which does not require to
build any model at all.

The linear temporal logic we use to express specifications, \csltl, is an
adaptation of the propositional \ltl\ logic to the concurrent constraint
framework.
This logic is also used as the basis of the abstract domain for a new
(abstract) semantics for the language.

In brief, our method is an \emph{extension} of abstract diagnosis for
\tccp\ \cite{CominiTV11absdiag} where the abstract domain $\Fdom$ is formed
by \csltl\ formulas.  We cannot use the original abstract diagnosis
framework of \cite{CominiTV11absdiag} since $\Fdom$ is not a complete
lattice.

The contributions of this work are the following:
\begin{itemize}
    
    \item 
    A new abstract semantics for \tccp{} programs based on \csltl{}
    formulas;
    
    \item 
    A novel and effective method to validate \csltl{} properties based on
    the ideas of abstract diagnosis.  This proposal intuitively consists in
    viewing $P$ as a formula transformer by means of an (abstract)
    immediate consequence operator $\FDd{P}{}$ which works on \csltl\
    formulas.  Then, to decide the validity of $\phi$, we just have to
    check if $\FDd{P}{\phi}$ (\ie\ the $P$-transformation of $\phi$)
    implies $\phi$;
    
    \item 
    An automatic decision procedure for \csltl{} properties that makes our
    method effective.
\end{itemize}

With our technique we can check, for instance, that, at a railway crossing
system, each time a train is approaching, the gate is down, or that
whenever a train has crossed, the gate is up.  When a property is non
valid, the method identifies the buggy \progrule.  Technical results of
Sections~\ref{sec:CLTLabstr} and~\ref{sec:abs-diag} can be found in
\cite{CominiTV14-techrep}.

\section{The small-step operational behavior of the \tccp\ language}
\label{sec:Sem}

The \tccp{} language \cite{deBoerGM99} is particularly suitable to specify
reactive and time critical systems.  As the other languages of the \ccp{}
paradigm \cite{Saraswat93}, it is parametric \wrt\ a \emph{cylindric
constraint system} which handles the data information of the program in
terms of constraints.  The computation progresses as the concurrent and
asynchronous activity of several agents that can accumulate information in
a \emph{store}, or query information from it.  Briefly, a cylindric
constraint system\footnote{See \cite{deBoerGM99,Saraswat93} for more
details on cylindric constraint systems.} is an algebraic structure $\CSys=
\langle \CSdom, \mathord{\CSord}, \CSmerge, \CSfalse, \CStrue,
\mathord{\Var}, \CShid{}{} \rangle$ composed of a set of constraints
$\CSdom$ such that $\cpo{\CSdom}{\CSord}$ is a complete algebraic lattice
where $\CSmerge$ is the $\lub$ operator and $\CSfalse$ and $\CStrue$ are
respectively the greatest and the least element of $\CSdom$; $\Var$ is a
denumerable set of variables and $\CShid{}$ existentially quantifies
variables over constraints.  The \emph{entailment} $\CSimp$ is the inverse
of $\CSord$.

Given a cylindric constraint system $\CSys$ and a set of process symbols
$\Pi$, the syntax of agents is given by the grammar:
\begin{align*}
    A ::= \askip \mid \atell{c} \mid A \parallel A \mid
    \ahiding{x}{A} \mid
    \textstyle{\sum_{i=1}^{n}\aask{c_{i}} A}
    \mid \anow{c}{A}{A}
    \mid p(\zseq{x})
\end{align*}
where $c$, $\seq{c}$ are finite constraints in $\CSys$; $p_{/m}\in\Psyms$
and $\zseq{x}$ denotes a generic tuple of $m$ variables.  A \tccp{} \query\
is an object of the form $\Qq{D}{A}$, where $A$ is an agent, called
\emph{initial agent}, and $D$ is a set of \emph{process declarations} of
the form $\mgrule{p}{x}{A}$ (for some agent $A$).  The notion of time is
introduced by defining a discrete and global clock.
\begin{figure}[t]
    \begin{minipage}{\textwidth}
        {\footnotesize
        {\setlength{\jot}{1.5ex}
        \begin{align*}
            & \sequent{}
            {\pair{\atell{c}}{d} \ra \pair{\askip}{c \CSmerge d}
            }
            {$d\neq\CSfalse$}{}
            & \sequent{}
            {\pair{\asumask{n}{c}{A}}{d} \ra \pair{A_{j}}{d} }
            {$j
            \in[1,n]$, $d \CSimp c_{j}$,
            $d\neq\CSfalse$}{}
            \\
            & \sequent{\pair{A}{d} \ra \pair{A'}{d'} }
            {\pair{\anow{c}{A}{B}}{d} \ra \pair{A'}{d'} }
            {$d\CSimp c$}{}
            & \sequent{ \pair{A}{d} \nra }
            { \pair{\anow{c}{A}{B}}{d} \ra
            \pair{A}{d} }
            {$d\CSimp c$, $d\neq\CSfalse$ }{}
            \\
            & \sequent{ \pair{B}{d} \ra \pair{B'}{d'} }{
            \pair{\anow{c}{A}{B}}{d} \ra \pair{B'}{d'} }{$d\CSnimp c$ }{}
            & \sequent{ \pair{B}{d} \nra }{ \pair{\anow{c}{A}{B}}{d} \ra
            \pair{B}{d} }{$d\CSnimp c$ }{}  
            \\
            & \sequent{ \pair{A}{d} \ra \pair{A'}{d'} \quad \pair{B}{d} \ra
            \pair{B'}{c'} }{ \pair{A\parallel B}{d} \ra \pair{A'\parallel
            B'}{d'\CSmerge c'} }{ }{}
            & \sequent{ \pair{A}{d} \ra \pair{A'}{d'} \quad \pair{B}{d} \nra }{
            \begin{array}{c}
                \pair{A\parallel B}{d} \ra \pair{A'\parallel B}{d'} \\
                \pair{B\parallel A}{d} \ra \pair{B \parallel A'}{d'}
            \end{array}
            }{ }{}
            \\
            & \sequent{ \pair{A}{l \CSmerge \CShid{x}{d}} \ra\pair{B}{l'} }{
            \pair{\ahiding[l]{x}{A}}{d} \ra \pair{\ahiding[l']{x}{B}}{d
            \CSmerge \CShid{x}{l'}} }{ }{}
            & \sequent{ }{ \pair{ \mgc{p}{x}{} }{d} \ra \pair{A}{d}
            }{\:$\mgrule{p}{x}{A} \in D$, $d\neq\CSfalse$ }{}
        \end{align*}
        }
        }
        \caption[The transition system for \tccp{}.]{The transition system for \tccp{}.\footnotemark{}}
        \label{fig:op_sem}
    \end{minipage}
\end{figure}
\footnotetext{The auxiliary agent $\ahiding[l]{x}{A}$ makes explicit the
local store $l$ of $A$.  This auxiliary agent is linked to the principal
hiding construct by setting the initial local store to $\CStrue$, thus
$\ahiding{x}{A} \dfn \ahiding[\CStrue]{x}{A}$.  }

The \emph{operational semantics} of \tccp{}, defined in \cite{deBoerGM99},
is formally described by a transition system $T=(\Conf, \ra)$.
Configurations in $\Conf$ are pairs $\pair{A}{c}$ representing the agent
$A$ to be executed in the current global store $c$.  The transition
relation ${\ra} \subseteq \Conf\times\Conf$ is the least relation
satisfying the rules of Figure~\ref{fig:op_sem}.  Each transition step takes
exactly one time-unit.
\begin{example}[Guiding example]\label{ex:gateController}
    Through the paper, we use as guiding example a part of the full
    specification of a railway crossing system introduced in
    \cite{AlpuenteGPV06}.  Let us call $\Dmst$ the following \tccp{}
    declaration: {\small
    \begin{align*}
        &\prule{\controller}{C,G}{ \ahiding{C', G'}{} \big( }
        \begin{aligned}[t]
            &\anow{(C = [\mathit{near}\mid \_])}{}{}\\
            &\quad
            \atell{C = [\textit{near}\mid C']}\aparallel{}{}
            \atell{G = [\textit{down}\mid G']}\aparallel{}{}
            \call{\controller}{C',G'}{} \\
            &\tagelse
            \begin{aligned}[t]
                &\anow{(C = [\mathit{out}\mid \_])}{}{}\\
                &\quad
                \atell{C = [\textit{out}\mid C']}\aparallel{}{}
                \atell{G = [\textit{up}\mid G']}\aparallel{}{}
                \call{\controller} {C',G'}{}\\
                &\tagelse \call{\controller}{C,G}{}  \big)
            \end{aligned}
        \end{aligned}
    \end{align*}
    }%
    Due to the monotonicity of the store, streams (written in a
    list-fashion way) are used to model \emph{imperative-style} variables
    \cite{deBoerGM99}.  The \controller\ process uses an \emph{input
    channel} $C$ (implemented as a stream) through which it receives
    signals from the environment (trains), and an \emph{output channel} $G$
    through which it sends orders to a gate process.  It checks the input
    channel for a \textit{near} signal (the guard in the first
    $\anow{}{}{}$ agent), in which case it sends (tells) the order
    \textit{down} through $G$, links the future values ($C'$) of the stream
    $C$ and restarts the check at the following time instant (recursive
    call $\call{\controller}{C',G'}{}$).  If the \emph{near} signal is not
    detected, then, the else branch looks for the \emph{out} signal and (if
    present) behaves dually to the first branch.  Finally, if no signal is
    detected at the current time instant (last else branch), then the
    process keeps checking from the following time instant.
\end{example}

In this work, we prove the correctness of our technique \wrt\ the
denotational concrete semantics of \cite{CominiTV13semTR}, which is
fully-abstract (correct and complete) \wrt\ the small-step operational
behavior of \tccp{}.  Also \csltl{} is interpreted over this
denotational model.  We thus introduce the most relevant aspects of
such semantics.

The denotational semantics of a \tccp{} program consists of a set of
\emph{conditional (timed) traces} that represent, in a compact way, all the
possible behaviors that the program can manifest when fed with an
\emph{input} (initial store).  Conditional traces can be seen as
hypothetical computations in which, for each time instant, we have a
condition representing the information that the global store has to satisfy
in order to proceed to the next time instant.  Briefly, a conditional trace
is a (possibly infinite) sequence $t_1\cdots t_n\cdots$ of
\emph{conditional states}, which can be of three forms:
\begin{description}
    \item[conditional store:] 
    a pair $\cs{\eta}{c}$, where $\eta$ is a \emph{condition} and
    $c\in\CSys$ a store;
    
    \item[stuttering:] 
    the construct $\stutt{C}$, with $C \subseteq
    \CSys\setminus\{\CStrue\}$;
    
    \item[end of a process:] 
    the construct $\ed$.
\end{description}
Intuitively, the conditional store $\cs{\eta}{c}$ means that, provided
condition $\eta$ is satisfied by the current store, the computation
proceeds so that in the following time instant, the store is $c$.  A
\emph{condition} $\eta$ is a pair $\eta=(\eta^{+},\eta^{-})$ where
$\eta^{+} \in \CSys$ and $\eta^{-} \in \wp(\CSys)$ are called positive and
negative condition, \resp.  The positive/negative condition represents
information that a given store must/must not entail, thus they have to be
consistent in the sense that $\forall c^-\in\eta^-$ $\eta^+\CSnimp c^-$.
The stuttering construct models the suspension of the computation when none
of the guards in a non-deterministic agent is satisfied.  $C$ is the set of
guards in the non-deterministic agent.  Conditional traces are monotone
(\ie\ for each $t_i=\cs{\eta_i}{c_i}$ and $t_j=\cs{\eta_j}{c_j}$ such that
$j\geq i$, $c_j \CSimp c_i$) and consistent (\ie\ each store in a trace
does not entail the negative conditions of the following conditional
state).

We denote the domain of \emph{conditional trace sets} as $\C$.  $\latticeC$
is a complete lattice, where $M_{1} \leqC M_{2} \Leftrightarrow \forall
r_{1}\in M_{1} \, \exists r_{2}\in M_{2} .\, r_{1}\ \text{is a prefix of}\
r_{2}$.  We define as $\seqHid{x}{r}$ the sequence resulting by removing
from $r\in\topC$ all the information about the variable $x$.  We
distinguish two special classes of conditional traces.  $r\in\topC$ is said
to be \emph{\closed} if the first condition is $\cond{\CStrue}{\emptyset}$
and, for each $t_i = \csC{\eta^+_i}{\eta^-_i}{c_i}$ and
$t_{i+1}=\csC{\eta^+_{i+1}}{\eta^-_{i+1}}{c_{i+1}}$, $c_i \CSimp
\eta^+_{i+1}$ (each store satisfies the successive condition).  Moreover,
$r$ is \emph{$x$-\closed} if $\seqHid{\Var\setminus \{x\}}{r}$ is \closed.
Thus, this definition demands that for \closed\ conditional traces, no
additional information (from other agents) is needed in order to
\emph{complete} the computation.\footnote{The set of all \closed\
conditional traces can be considered as a generalization (using conditional
states in place of stores) of the traditional strongest postcondition for
semantics.}

The semantics definition is based on a semantics evaluation function
$\Aa{A}{\I[]}$ \cite{CominiTV13semTR} which, given an agent $A$ and an
interpretation $\I[]$, builds the conditional traces associated to $A$.
The interpretation $\I[]$ is a function which associates to each process
symbol a set of conditional traces ``modulo variance''.  The semantics for
a set of process declarations $D$ is the fixpoint $\F[]{\P} \dfn
\lfp(\Dd{\P}{})$ of the continuous immediate consequences operator
$\Dd{\P}{\I[]} (\mgc{p}{x}{}) \dfn \lubC{_{\mgrule{p}{x}{A} \in D}
\Aa{A}{\I[]}} {}$.  Proof of full abstraction \wrt\ the operational
behavior of \tccp{} is given in \cite{CominiTV13semTR}.
\begin{figure}[tp]
    \centering{
    \begin{tikzpicture}[scale=0.5]
        
        \draw (0,0) coordinate (root);
        \draw (root)+(-7.4,-0.5) node (near1) {$\csC{c_{\mathit{near}}}{\emptyset}{c_{\mathit{near}}' \wedge c_{\mathit{down}}}$};
        \draw (root)+(0,-1.1) node (out1) {$\csC{c_{\mathit{out}}}{\{c_{\mathit{near}\}}}{c_{\mathit{out}}' \wedge c_{\mathit{up}}}$};
        \draw (root)+(7.2,-0.5) node (rec1) {$\csC{\CStrue}{\{c_{\mathit{near}}, c_{\mathit{out}}\}}{\CStrue}$};
        \draw[-latex] (root) -- (near1);
        \draw[-latex] (root) -- (out1);
        \draw[-latex] (root) -- (rec1);
        
        \draw (near1.south)+(4,-2.3) coordinate (con13);
        \draw (near1.south)+(-4,-2.3) coordinate (con14);
        \draw (near1.south) -- (con13) -- (con14) -- cycle;
        \draw (near1.south)+(0,-1.4) node (con1call)
        {\small $
        \begin{aligned} 
            &\quad\qquad\seqHid{C',G'}{}\\
            &\I[](\call{\controller}{C',G'}{})
        \end{aligned}$};
        \draw (out1.south)+(4,-2.3) coordinate (con23);
        \draw (out1.south)+(-4,-2.3) coordinate (con24);
        \draw (out1.south) -- (con23) -- (con24) -- cycle;
        \draw (out1.south)+(0,-1.4) node (con2call)
        { \small $\begin{aligned} &\quad\qquad\seqHid{C',G'}{}\\ &\I[](\call{\controller} {C',G'}{})\end{aligned}$};
        \draw (rec1.south)+(-4,-1.8) coordinate (con33);
        \draw (rec1.south)+(4,-1.8) coordinate (con34);
        \draw (rec1.south) -- (con33) -- (con34) -- cycle;
        \draw (rec1.south)+(0,-1.4) node (con3call) {\small$\I[](\call{\controller} {C,G}{})$};
    \end{tikzpicture}
    }
    \caption{Tree representation of $\Dd{\Drc}{\I[]} (\call{\controller} {C,G}{})$  of
    \smartref{ex:TrainSmallStep}.}
    \label{fig:controllerSmallStep}
\end{figure}  
\begin{example}\label{ex:TrainSmallStep}
    Consider the \progrule{} $\Dmst$ of \smartref{ex:gateController}.
    Given an interpretation $\I[]$, the semantics of
    $\call{\controller}{C,G}{}$ is graphically represented in
    Figure~\ref{fig:controllerSmallStep}, where we have used some shortcuts
    for characteristic constraints.  Namely, $c_{\mathit{near}} \dfn (C =
    [\mathit{near} \mid \_])$, $c_{\mathit{near}}' \dfn \CShid{C'}{(C =
    [\mathit{near} \mid C'])}$, $c_{\mathit{down}} \dfn \CShid{G'} {(G =
    [\mathit{down} \mid G'])}$, $c_{\mathit{out}} \dfn (C = [\mathit{out}
    \mid \_])$, $c_{\mathit{out}}' \dfn \CShid{C'} {(C = [\mathit{out} \mid
    C'])}$, $c_{\mathit{up}} \dfn \CShid{G'} {(G = [\mathit{up} \mid
    G'])}$.
    
    The branch on the left represents the computation when a \emph{near}
    signal arrives.  The first conditional state requires that
    $c_{\mathit{near}}$ holds, thus the constraints $c_\mathit{near}'$ and
    $c_\mathit{down}$ are \emph{concurrently} added to the store during
    that computational step.  A recursive call is also concurrently
    invoked.  Process calls do not modify the store when invoked, but they
    affect the store from the following time instant, which is graphically
    represented by the triangle labeled with the interpretation of the
    process.  The branch in the middle is taken only if $c_\mathit{out}$ is
    entailed and $c_\mathit{near}$ is not entailed by the initial store (it
    occurs in the negative condition of the first conditional state in that
    branch).  Finally, the branch on the right represents the case when
    both $c_\mathit{near}$ and $c_\mathit{out}$ are not entailed by the
    initial store.
\end{example}

\section{Abstract semantics for \tccp\ over \csltl\ formulas} 
\label{sec:CLTLabstr}

In this section, we present a novel abstract semantics over formulas that
approximates the small-step semantics described in \smartref{sec:Sem} and,
therefore, the small-step operational behavior of a \tccp\ program.  To
this end, we first define an abstract domain of logic formulas which is a
variation of the classical Linear Temporal Logic \cite{MannaP92}.
Following \cite{PalamidessiV-CP2001,deBoerGM01,deBoerGM02,Valencia05}, the
idea is to replace atomic propositions by constraints of the underlying
constraint system.
\begin{definition}[\csltl{} formulas] \label{def:tf}
    Given a cylindric constraint system $\CSys$, $c\in\CSys$ and
    $x\in\Var$, formulas of the \emph{Constraint System Linear Temporal
    Logic} over $\CSys$ are:
    \begin{equation*}
        \phi ::= \Ltrue \mid \Lfalse \mid c \mid \Lneg{\phi} \mid
        \Lconj{\phi}{\phi} \mid \Lhid{x}{\phi} \mid \Lnext{\phi} \mid
        \Luntil{\phi}{\phi}.
    \end{equation*}
    $\csltl[\CSys]$ is the set of all temporal formulas over $\CSys$ (we
    omit $\CSys$ if clear from the context).
\end{definition}
$\Ltrue$, $\Lfalse$, $\Lneg{}$, $\Lconj{}{{}}$, $\Lnext{\phi}$,
$\Luntil{\phi_1}{\phi_2}$ have the classical logical meaning.  The atomic
formula $c\in\CSys$ states that $c$ has to be entailed by the current
store.  $\Lhid{x}{\phi}$ is the existential quantification over the set of
variables $\Var$.  As usual, we use $\Ldisj{\phi_1}{\phi_2}$ as a shorthand
for $\Lconj{\Lneg{\phi_1}}{\Lneg{\phi_2}}$; $\Limpl{\phi_1}{\phi_2}$ for
$\Ldisj{\Lneg{\phi_1}}{\phi_2}$; $\Liff{\phi_1}{\phi_2}$ for
$\Lconj{\Limpl{\phi_1}{\phi_2}}{\Limpl{\phi_2}{\phi_1}}$;
$\Leventually{\phi}$ for $\Luntil{\Ltrue}{\phi}$ and $\Lalways{\phi}$ for
$\Lneg{\Leventually{\Lneg{\phi}}}$.  A \emph{constraint formula} is an
atomic formula $c$ or its negation $\Lneg{c}$.  Formulas $\Lnext{\phi}$ and
$\Lneg{\Lnext{\phi}}$ are called \emph{next} formulas.  Constraint and next
formulas are said to be \emph{elementary} formulas.  Finally, formulas of
the form $\Luntil{\phi_1}{\phi_2}$, $\Leventually{\phi}$ or
$\Lneg{(\Lalways{\phi})}$ are called \emph{eventualities}.

We define the abstract domain $\Fdom \dfn \csltl/_{\Liff{}{}}$ (\ie{} the
domain formed by \csltl{} formulas modulo logical equivalence) ordered by 
$\Limpl{}{}$.
The algebraic lattice $\latticeF$ is not complete, since both $\glbF{}{}$
and $\lubF{}{}$ always exist just for finite sets of formulas.

The semantics of a temporal formula is typically defined in terms of an
infinite sequence of states which validates it.  Here we use conditional
traces instead.  As usually done in the context of temporal logics, we
define the satisfaction relation $\asat{}{}$ only for infinite conditional
traces.  We implicitly transform finite traces (which end in $\ed$) by
replicating the last store infinite times.
\begin{definition}
    \label{def:FtoM}
    \label{def:asat}
    The \emph{semantics} of $\phi\in\Fdom$ is given by function $\FtoM{}
    \colon \Fdom \rightarrow \C$ defined as
    $\FtoM{\phi} \dfn \lubC{ \set*{ r\in \topC }{ \asat{r}{\phi} } }{}$,
    where, for each $\phi, \phi_1, \phi_2 \in \csltl$, $c \in \CSys$ and
    $r\in\topC$, satisfaction relation $\asat{}{}$ is defined as:
    \begin{subequations}
        \small
        \label{eq:asat_rel}
        \begin{align}
            &\asat{r}{\Ltrue} \quad\text{and}\quad \notsat{r}{\Lfalse}\\
            &\asat{\csC{\eta^+}{\eta^-}{d} \cdot r'}{c} &&\text{iff }
            \eta^+ \CSimp c
            \label{eq:asatKnowCS}\\
            &\asat{\stutt{\eta^-} \cdot r'}{c} &&\text{iff } \forall
            d^{-}\in \eta^- .\, c \CSnimp d^{-} \text{ and } \asat{r'}{c}
            \label{eq:asatKnowST}\\
            &\asat{r}{\Lneg{\phi}} &&\text{iff } \notasat{r}{\phi}
            \label{eq:asatNeg} \\
            &\asat{r}{\Lconj{\phi_1}{\phi_2}} &&\text{iff } \asat{r}{\phi_1}
            \text{ and } \asat{r}{\phi_2} \label{eq:asatConj} \\
            & \asat{r}{\Lnext{\phi}} &&\text{iff}\ \asat{r^1}{\phi} 
            ~\footnotemark
            \label{eq:asatNext} \\
            & \asat{r}{\Luntil{\phi_1}{\phi_2}} &&\text{iff }            
            \exists i\geq
            1 .\, \forall j < i .\ \asat{r^i}{\phi_2}\ \text{and}\ \asat{r^j}{\phi_1}
            \label{eq:asatUntil} \\
            &\asat{r}{\Lhid{x}{\phi}} 
            \quad\quad\quad\; \makebox[0pt][l]{iff exists $r'$ \st\
            $\seqHid{x}{r'}=\seqHid{x}{r}$, $r'$ $x$-\closed\ and
            $\asat{r'}{\phi}$}
            \label{eq:asatHid}
        \end{align}
    \end{subequations}%
    \footnotetext{$r^{k}$ denotes the sub-sequence of $r$ starting from
    state $k$.  }%
    We say that $\phi\in\Fdom$ is a \emph{sound approximation} of $R \in \C$ if $R
    \leqC \FtoM{\phi}$.
    $\phi$ is said to be \emph{satisfiable} if there exists
    $r\in\topC$ such that $\asat{r}{\phi}$, while it is
    \emph{valid} if, for all $r\in\topC$, $\asat{r}{\phi}$.
\end{definition}

All the cases are fairly standard except \eqref{eq:asatKnowCS} and
\eqref{eq:asatKnowST}.
The conditional trace $r=\csC{\eta^+}{\eta^-}{d} \cdot r'$ prescribes that
$\eta^+$ is entailed by the current store, thus $r$ models all the
constraint formulas $c$ such that $\eta^+ \CSimp c$.  We have to note that,
by the monotonicity of the store of \tccp{} computations, the positive
conditions in conditional traces contains all the information previously
added in the constraint store.
Furthermore, by the definition of condition, since $\eta^+$ cannot be in
contradiction with $\eta^-$, it holds that neither $c$ is in contradiction
with $\eta^-$.  Thus, the conditional trace $\stutt{\eta^-} \cdot r'$
models all the constraint formulas $c$ that are not in contradiction with
the set $\eta^-$ and such that $c$ holds in the continuation $r'$ by
monotonicity.
\begin{lemma}\label{lem:FtoM}
    The function $\FtoM{}$ is monotonic, injective
    and $\glbC{}{}$-distributive.
\end{lemma}

\subsection{\csltl\ Abstract Semantics}

The technical core of our semantics definition is the \csltl\ agent
semantics evaluation function $\FAa{A}{}$ which, given an agent $A$ and an
interpretation $\FI$ (for the process symbols of $A$), builds a \csltl\
formula which is a sound approximation of the (concrete) behavior of $A$.
In the sequel, we denote by $\Agents$ the set of agents and $\Progs$ the
set of sets of process declarations built on signature $\Psyms$ and
constraint system $\CSys$.
\begin{definition}
    \label{def:interpC}
    Let $\MGC \dfn \{ p(\zseq{x}) \mid p\in \Pi$, $\zseq{x}$ are distinct
    variables$\,\}$.
    An \emph{$\Fdom$-interpretation} is a function $\MGC \to \Fdom$ modulo
    variance\footnote{\ie\ a family of elements of $\Fdom$, indexed by
    $\MGC$, modulo variance.  }.  Two functions $I,J \colon \MGC \to \Fdom$
    are \emph{variants}
    if for each $\pi \in \MGC$ there exists a renaming $\rho$ such that
    $(I\pi)\rho =J(\pi\rho)$.
    The semantic domain $\interpF$ is the set of all
    $\Fdom$-interpretations ordered by the point-wise extension of $\leqF$.
\end{definition}    
\begin{definition}[\csltl\ Semantics]
    \label{def:semFAa} 
    Given $A\in\Agents$ and $\FI \in \interpF$, we define the \emph{\csltl\
    semantics evaluation} $\FAa{A}{\FI}$ by structural induction as
    follows.
    \begin{align*}
        & \FAa{\askip}{\FI} \dfn \CStrue
        \qquad\qquad\qquad\qquad \FAa{A_1 \parallel A_2}{\FI} \dfn \Lconj{\FAa{A_1}{\FI}}{\FAa{A_2}{\FI}}
        \\& \FAa{\atell{c}}{\FI} \dfn \Lnext{c}
        \qquad \FAa{\ahiding{x}{A}}{\FI} \dfn \Lhid{x}{\FAa{A}{\FI}}
        \qquad \FAa{ \mgc{p}{x}{} }{\FI} \dfn \Lnext{\FI( \mgc{p}{x}{} )}
        \\
        & \FAa{\textstyle{\asumask{n}{c}{A}}}{\FI} \dfn
        \Ldisj{\Lalways{(\Lconj{_{i=1}^{n} \Lneg{c_i}}{})}}{{}}
        {\Big(\Luntil{\big(\Lconj{_{i=1}^{n}
        \Lneg{c_i}}{}\big)}{\Ldisj{_{i=1}^{n}\big(\Lconj{c_i}{\Lnext{\FAa{A_i}{\FI}}}\big)}{}}\Big)}
        \\
        & \FAa{\anow{c}{A_1}{A_2}}{\FI} \dfn
        \Ldisj{(\Lconj{c}{\FAa{A_1}{\FI}})}{(\Lconj{\Lneg{c}}{\FAa{A_2}{\FI}})}
    \end{align*}
    Given $\P\in\Progs$ we define the immediate consequence operator
    $\FDd{\P}{} \colon \interpF \to \interpF$ as
    \begin{equation*}
        \FDd{\P}{\FI}( \mgc{p}{x}{} ) \dfn \lubF{ \set*{ \FAa{A}{\FI} }{
        \mgrule{p}{x}{A} \in D} }{}
    \end{equation*}
\end{definition}

We have that $\FAa{}{}$ is a sound approximation of $\Aa{}{}$ and
$\FDd{}{}$ is a sound approximation of $\Dd{}{}$.
\begin{theorem}[Correctness of $\FAa{}{}$ and $\FDd{}{}$]
    \label{th:FAa_FDd_soundness}
    Let $A\in\Agents$, $\P\in\Progs$ and $\FI \in \interpF$.  Then,
    $\Aa{A}{\FtoM{\FI}} \leqC \FtoM{\FAa{A}{\FI}}$ and
    $\Dd{\P}{\FtoM{\FI}} \leqC \FtoM{\FDd{\P}{\FI}}$.
\end{theorem}
\begin{example}\label{ex:TrainAbsSem}
    Consider the \progrule{} $\Dmst$ of \smartref{ex:gateController} and
    let us use $\Lnextnum{n}{}$ to abbreviate the repetition of $\Lnext$
    $n$-times.
    Given $\FI \in \interpF$, with \smartref{def:semFAa} we compute
    \begin{align*}
        & \phicont{\FI} \dfn \FDd{ \Drc }{\FI} (\call{\controller}
        {C,G}{})) =
        \Ldisj{ \Ldisj{\phi_{\mathit{near}}(\FI)}
        {\phi_{\mathit{out}}(\FI)} }{\phi_{\mathit{cwait}}(\FI)} 
    \end{align*}
    where 
    {\small
    \begin{align*}
        &\phi_{\mathit{near}}(\FI) =
        \Lhid{C',G'}{
        \begin{aligned}[t]\big(
            &\Lconj{C = [\textit{near} \mid \_]}{{}} \Lconj{
            \Lnext{ C = [\textit{near} \mid C']}}{{}}
            \Lconj{\Lnext{G = [\textit{down} \mid G']}}
            {\Lnext{\FI(\call{\controller}{C',G'}{})}} \big)
        \end{aligned}} \\
        &\phi_{\mathit{out}}(\FI) = \Lhid{C',G'}{
        \begin{aligned}[t]
            \big( &\Lconj{\Lneg{(C =
            [\textit{near} \mid \_])} } { \Lconj{\Lnext{ C = [\textit{out} \mid C']}}{{}} } \\*
            &\Lconj{C=[\textit{out} \mid \_]}{
            \Lconj{\Lnext{G = [\textit{up} \mid G']}}{\Lnext{\FI(\call{\controller}{C',G'}{})}} } \big)
        \end{aligned}}\\
        &\phi_{\mathit{cwait}}(\FI) =\Lconj{\Lneg{(C = [\textit{near} \mid
        \_])}} { \Lconj {\Lneg{(C = [\textit{out} \mid \_])}} { {} }}
        \Lnext{\FI(\call{\controller}{C,G}{})}
    \end{align*}
    }%
    The three disjuncts of $\phicont{\FI}$ match the three possible
    behaviors of $\call{\controller} {C,G}{}$: when signal $\mathit{near}$
    is emitted by the train, when $\mathit{out}$ is emitted, and when no
    signal arrives.
\end{example}

\section{Abstract diagnosis of \tccp\ with \csltl\ formulas}
\label{sec:abs-diag}

Since $\Fdom$ is not a complete lattice, it is impossible to find for the
function $\FtoM{}$ an adjoint function $\alpha$ which forms a Galois
Connection $\pair{\alpha}{\gamma}$, and therefore we cannot use the
abstract diagnosis framework for \tccp\ defined in
\cite{CominiTV11absdiag}.  Thus, we propose in this section a new weaker
version of abstract diagnosis that works on $\Fdom$ \footnote{Actually, the
proposal is defined using just $\FtoM{}$ only for the sake of simplicity.
It could easily be defined parametrically \wrt\ a suitable family of
concretization functions.  }.

Given a \program\ $\P$ and $\SF \in \interpF$, which is the specification
of the abstract intended behavior of $D$ over $\Fdom$, we say that
\begin{enumerate}
    
    \item\label{pt:COR:def:Correct}
    $\P$ is (abstractly) \emph{partially correct}\index{partially correct}
    \wrt\ $\SF$ if $\F[]{\P} \leqC \FtoM{\SF}$.
    
    \item\label{pt:COM:def:Correct}
    $\P$ is (abstractly) \emph{complete}\index{complete} \wrt\ $\SF$ if
    $\FtoM{\SF} \leqC \F[]{\P}$.
\end{enumerate}
The differences between $\F[]{\P}$ and $\FtoM{\SF}$ are usually called
\emph{symptoms}.  Many of the symptoms are just a consequence of some
``originating'' ones, those which are the direct consequence of errors.
The \emph{abstract diagnosis} determines exactly the ``originating''
symptoms and, in the case of incorrectness, the faulty \progrules\ in $\P$.
This is captured by the definitions of \emph{abstractly incorrect
\progrule} and \emph{abstract uncovered element}:\footnote{It is worth
noticing that although the notions defined in this section are similar to
those defined for the standard approach, the formal definitions and proofs
are different due to the weaker framework.}
\begin{definition}
    \label{def:ab.incorclau}
    \label{def:ab.uncovered}
    
    Let $\P\in\Progs$, $R$ a \progrule\ for \deffun[long] $p$,
    $\phi_{t}\in\Fdom$ and $\SF \in \interpF$.
    \begin{itemize}
        \item $R$ is \emph{abstractly incorrect} \wrt\ $\SF$ (on testimony
        $\phi_{t}$) if $\phi_{t} \leqF \FDd{\{R\}}{\SF}(\mgc{p}{x}{})$ and
        $\glbF{ \phi_{t} }{ {} }$ $\SF(\mgc{p}{x}{}) = \botF$.
        
        \item $\phi_{t}$ is an \emph{uncovered element} for $\mgc{p}{x}{}$
        \wrt\ $\SF$ if $\phi_{t} \leqF \SF(\mgc{p}{x}{})$ and
        $\glbF{\phi_{t}}{\FDd{\P}{\SF}(\mgc{p}{x}{})} = \botF$.
    \end{itemize}
\end{definition}
Informally, $R$ is abstractly incorrect if it derives a wrong abstract
element $\phi_{t}$ from the intended semantics.  Dually, $\phi_{t}$ is
uncovered if the declarations cannot derive it from the intended semantics.
\begin{theorem}
    \label{th:ab.corr-compl}
    Let $\P\in\Progs$ and $\SF \in \interpF$. (1)
    If there are no abstractly incorrect
    process declarations in $\P$ (\ie\ $\FDd{\P}{\SF} \leqF \SF$),
    then $\P$ is partially correct \wrt\ $\SF$.
    (2) If $\P$ is partially correct \wrt\ $\SF$ and $\P$ has abstract
    uncovered elements then $\P$ is not complete.
\end{theorem}
Absence of abstractly incorrect declarations is a sufficient condition for
partial correctness, but it is not necessary.  Because of the
approximation, it can happen that a (concretely) correct declaration is
abstractly incorrect.  Hence, abstract incorrect declarations are in
general just a warning about a possible source of errors.  However, an
abstract correct declaration cannot contain an error; thus, no (manual)
inspection is needed for declarations which are not abstractly incorrect.
Moreover, as shown by the following theorem, all concrete errors---that are
``visible''---are indeed detected, as they lead to an abstract
incorrectness or abstract uncovered.  Intuitively, a concrete error is
\emph{visible} if we can express a formula $\phi$ whose concretization
reveals the error (\ie\ if the logic is expressive enough).
\begin{theorem}
    \label{th:abs-conc}
    
    Let $R$ be a process declaration for $\mgc{p}{x}{}$, $\Sz[]$ a concrete
    specification and $\SF$ a sound approximation for $\Sz[]$ (\ie\ $\Sz[]
    \leqC \FtoM{\SF}$).
    (1) If $\Tp[]{\{R\}}{\Sz[]} \nleqC \FtoM{\SF}$ and it exists $\phi_t$
    such that $\FtoM{\phi_t} \leqC \Tp[]{\{R\}}{\Sz[]}(\mgc{p}{x}{})$
    and $\Lconj{\phi_t}{\SF(\mgc{p}{x}{})} = \Lfalse$, then $R$ is
    abstractly incorrect \wrt\ $\SF$ (on testimony $\phi_t$).
    (2) If there exists an abstract uncovered element $\phi$ \wrt\
    $\SF$, then there exists $r\in\FtoM{\phi}$ such that $r \nin
    \Tp[]{\{R\}}{\Sz[]}(\mgc{p}{x}{})$.
\end{theorem}
Point~2 says that the concrete error has an abstract symptom which is not
hidden by the approximation on $\SF$ and, moreover, there exists a formula
$\phi_t$ which can express it.

In the following examples, we borrow from \cite{AlpuenteGPV06} the notation
for \emph{last entailed value} of a stream: $X\dot{=}c$ holds if the last
instantiated value in the stream $X$ is $c$.
\begin{example}\label{ex:controllDiag-validProp} 
    We verify (for \smartref{ex:gateController}) that each time a
    \emph{near} signal arrives from a train, the order \emph{down} is sent
    to a gate process.\footnote{A more interesting property, namely that,
    in addition, the gate is eventually down, is verified in
    \cite{CominiTV14-techrep}.  Here we have simplified the property due to
    space limitations.} To model this property, we define the specification
    (of the property) $\SF_{down}$ as
    \begin{align*}
        &\phi_{\mathit{ordersent}} \dfn \SF_{\mathit{down}}(\call{\controller}{C,G}{}) 
        \dfn \Lalways{(\Limpl{C \dot{=} \mathit{near}} {\Leventually{(G
        \dot{=} \mathit{down}})})}\\[-5ex]
    \end{align*}
    To check whether the program implies the specification
    ($\FDd{\Drc}{\SF_{\mathit{down}}} \leqF \SF_{\mathit{down}}$) we have
    to check if $\phicont{\SF_{\mathit{down}}} \leqF
    \phi_{\mathit{ordersent}}$ (where $\phicont{\cdot}$ is defined in
    \smartref{ex:TrainAbsSem}).  Each of the three disjuncts of
    $\phicont{\SF_{\mathit{down}}}$ implies $\phi_\mathit{ordersent}$.
    Thus, by \smartref{th:ab.corr-compl}, $\Dmst$ is partially correct
    \wrt\ $\SF_{\mathit{down}}$.
\end{example}

When the check of a \progrule\ $R$ against a specification $S$ fails, our
method reports that $R$ is not partially correct \wrt\ $S$.  If this
occurs, the formula testimony for the possible incorrectness gives useful
information to fix the \progrule\ or check whether it corresponds to a
false positive.
\begin{example}\label{ex:controllDiag-buggyDecl}
    Now we show how our technique detects an error in a buggy set of
    declarations.  We remove instruction $\atell{G = [\textit{up}\mid G']}$
    in the \progrule\ $\Dmst$ (of \smartref{ex:gateController}).
    To avoid misunderstandings, we call the modified process
    $\controller'$ and let $R$ be the new \progrule.
    
    We aim to verify that the order \emph{up} is sent whenever the
    signal \emph{out} is received: 
    \begin{align*}
        \phi \dfn \SF_{\mathit{up}}(\call{\controller'}{C,G}{}) \dfn 
        \Lalways{(\Limpl{(C \dot{=} \mathit{out})} {\Leventually{(G \dot{=}
        \mathit{up}})})}
    \end{align*}
    We need to compute the (one step) semantics for the (buggy version
    of the) process:
    \begin{equation*}
        \phi' \dfn \FDd{ \{R\} }{\SF_\mathit{up}} (\call{\controller'}{C,G}{}) =
        \Ldisj{ \Ldisj{\phi'_{near}} {\phi'_{out}}
        }{\phi'_{\mathit{cwait}}}
    \end{equation*}
    {\normalsize where}{\small
    \begin{align*}
        \phi'_{\mathit{near}} &\dfn \Lhid{C',G'}{
        \begin{aligned}[t]
            \big( &\Lconj{C = [\textit{near} \mid \_]}{\Lnext{ \Lconj{C = [\textit{near} \mid C']}}}
            \Lconj{\Lnext{G = [\textit{down} \mid G']}} {
            \Lnext{\SF_{\mathit{up}}(\call{\controller'}{C',G'}{}} } )\big)
        \end{aligned}
        }\\
        \phi'_{\mathit{out}} &\dfn \Lhid{C',G'}{
        \begin{aligned}[t]
            &\big( \Lconj{\Lneg{( C = [\textit{near} \mid \_] )}} {C =
            [\textit{out} \mid \_] } \Lconj{ }{ }
            \Lconj{ \Lnext{ (C = [\textit{out} \mid C'] }}{{}}
            \Lnext{\SF_{\mathit{up}}(\call{\controller'}{C',G'}{})})\big)
        \end{aligned}
        }\\
        \phi'_{\mathit{cwait}} &\dfn \Lconj{\Lneg{( C = [\textit{near} \mid
        \_] )}} { \Lconj {\Lneg{( C = [\textit{out} \mid \_] )}}
        {\Lnext{\SF_{\mathit{up}}(\call{\controller'}{C,G}{})}}}
    \end{align*}
    }%
    We detect an incorrectness of $R$ (in $\controller'$ process)
    \wrt\ $\SF_{\mathit{up}}$ on testimony $\phi'_\mathit{out}$ since
    $\phi'_{\mathit{out}} \leqF \phi'$ and $\glbF{ \phi'_{\mathit{out}}
    }{ \phi } = \botF$. 
    The testimony suggests that on channel $C$ we have $\mathit{out}$
    signal but we do not see the corresponding $\mathit{up}$
    signal on channel $G$.
\end{example}

Our technique behaves negatively for \programs\ $\P$ where $\FDd{\P}{}$ has
more than one fixpoint.  This happens with programs with loops that do not
produce contributes at all (which are in some sense non meaningful
programs).  In such situations, we can have that the actual behavior does
not model a specification $\SF$ which is a non-least fixpoint of
$\FDd{\P}{}$, but, since $\SF$ is a fixpoint, we do not detect the
abstractly incorrect declaration, as shown by the following example.
\begin{example}[Pathological cases]\label{ex:event_loop}
    Let $\P[p] \dfn \{
    \prule{q}{y}{\anow{y=1}{\apcall{q}{y}}{\apcall{q}{y}}} \}$ and $\SF_{p}
    (\apcall{q}{y}) \dfn \Leventually{(y=1)}$ be the specification.  Then,
    we compute $\FDd{ D_{p} }{\SF_{p}} ( \apcall{q}{y} ) =
    \Ldisj{(\Lconj{y=1}{\Leventually{y=1}})}
    {(\Lconj{\Lneg{y=1}}{\Leventually{y=1}})}$.  We can see that
    $\Limpl{\FDd{ D_{p} }{\SF_{p}}} {\Leventually{(y=1)}}$, thus $D_{p}$ is
    partially correct \wrt\ $\SF_{p}$.  However, $y=1$ is not explicitly
    added by the process.
\end{example} 

Note that, if $\SF (\mgc{p}{x}{})$ is assumed to hold for
each process $\mgc{p}{x}{}$ defined in $\D$ and $\Limpl{\FDd{ \D
}{\SF}} {{\SF}}$, then $\F[]{\P}$ satisfies $\SF$.

\subsection{An automatic decision procedure for \csltl}\label{sec:decision}

In order to make our abstract diagnosis approach effective, we have defined
an automatic decision procedure to check the validity of the formulas
involved in \smartref{def:ab.incorclau} (of the form $\Limpl{\psi}{\phi}$
with $\phi = \SF(\mgc{p}{x}{})$ and $\psi = \FDd{ \D }{\SF}
(\mgc{p}{x}{})$).  We adapt to \csltl\ the tableau construction for
Propositional \ltl\ of \cite{GaintzarainHLN08,GaintzarainHLNO09}.
\cite{CominiTV13-WLPE} contains a preliminary version of the method.

Intuitively, a tableau consists of a tree whose nodes are labeled
with sets of formulas.  The root is labeled with the set of formulas which
has to be checked for satisfiability.  Branches are built according to
rules defined on the syntax of formulas (see \smartref{tab:alpha_rules}
defining $\alpha$ and $\beta$ formulas). The basic idea is that a
formula from a node is selected and, depending on its form, a rule of
\smartref{tab:alpha_rules} is applied. $\beta$ formulas generate a
bifurcation on the tree and there are specific rules for next and
existential quantification formulas.

If all branches of the tree are \emph{closed}
(\smartref{def:node-inconsistent}), then the formula has no models.
Otherwise, we can obtain a model from the \emph{open} branches.
\begin{table*}
    \caption{$\alpha$- and $\beta$-formulas rules.}\label{tab:alpha_rules}\label{fig:beta_rules}
    \medskip
    \begin{minipage}{0.27\textwidth}
        {\small
        \begin{tabular}{  c  c  c  }
            \noalign{\vspace{-1ex}}
            & $\alpha$ \vspace{-.5ex}& $\Arule{\alpha}$\vspace{-.5ex}\\ \hline\hline
            \noalign{\vspace{-1ex}}
            \bf R1\ \setlabel{R1}\label{rule:neg} \!\!\!\!\!\vspace{-.4ex}& $\Lneg{\Lneg{\phi}}$ \vspace{-.4ex}& $\set{\phi}{}$ \vspace{-.4ex}\\ \hline
            \noalign{\vspace{-1ex}}
            \bf R2\setlabel{R2}\label{rule:conj} \!\!\!\!\!\vspace{-.4ex}& $\Lconj{\phi_1}{\phi_2}$ \vspace{-.4ex}& $\set{\phi_1,\phi_2}{}$ \vspace{-.4ex}\\ \hline
            \noalign{\vspace{-1ex}}
            \bf R3\setlabel{R3}\label{rule:negnext} \!\!\!\!\!\vspace{-.4ex}& $\Lneg{\Lnext{\phi}}$ \vspace{-.4ex}& $\set{\Lnext{\Lneg{\phi}}}{}$ \vspace{-.4ex}
        \end{tabular}
        }
    \end{minipage}
    \vrule width 1pt\,
    \begin{minipage}{0.68\textwidth}
        \begin{tabular}{  c  c  c  c  }
            \noalign{\vspace{-1ex}}
            & $\beta$ {\vspace{-.5ex}}& $\Brulel{\beta}$ {\vspace{-.5ex}}& $\Bruler{\beta}${\vspace{-.5ex}}\\ \hline\hline
            \noalign{\vspace{-1ex}}
            \bf R4 \setlabel{R4}\label{rule:negconj} \!\!\!\!\!{\vspace{-.3ex}}& $\Lneg{(\Lconj{\phi_1}{\phi_2})}$ {\vspace{-.3ex}}& $\set{\Lneg{\phi_1}}{}$ {\vspace{-.3ex}}& $\set{\Lneg{\phi_2}}{}$ {\vspace{-.3ex}}\\ \hline
            \noalign{\vspace{-1ex}}
            \bf R5 \setlabel{R5}\label{rule:neguntil} \!\!\!\!\!{\vspace{-.3ex}}& $\Lneg{(\Luntil{\phi_1}{\phi_2})}$ {\vspace{-.3ex}}& $\set{\Lneg{\phi_1},\Lneg{\phi_2}}{}$
            {\vspace{-.3ex}}& $\set{\phi_1,\Lneg{\phi_2},\Lneg{\Lnext{(\Luntil{\phi_1}{\phi_2})}}}{}$ {\vspace{-.3ex}}\\ \hline
            \noalign{\vspace{-1ex}}
            \bf R6 \setlabel{R6}\label{rule:dist_until} \!\!\!\!\!{\vspace{-.4ex}}& $\Luntil{\phi_1}{\phi_2}$ {\vspace{-.4ex}}&
            $\set{\phi_2}{}$ {\vspace{-.4ex}}& $\set{\phi_1,\Lneg{\phi_2},\Lnext{(\Luntil{(\Lconj{\cntx{\Gamma}}{\phi_1})}{\phi_2})}}{}$ {\vspace{-.4ex}}
        \end{tabular}
    \end{minipage}
\end{table*}
\begin{definition}[\csltl{} tableau]\label{def:tableau}
    A \csltl{} tableau for a finite set of formulas $\Phi$ is a tuple
    $\Tname = \Ttabl$ such that:
    \begin{enumerate}
        
        \item $\Tnodes$ is a finite non-empty set of nodes;
        
        \item $\Tinit \in \Tnodes$ is the initial node;
        
        \item $\Tlabel{}: \Tnodes \rightarrow \wp(\csltl)$ is the labeling
        function that associates to each node the formulas which are true
        in that node; the initial node is labeled with $\Phi$;
        
        \item
        $\Tbranches$ is the set of branches such that exactly one of the
        following points holds for every branch
        $b = n_0,
        \dots,n_i,n_{i+1},\dots,n_l \in \Tbranches$ and every $0 \leq i < l$:
        {\small
        \begin{enumerate}
            
            \item\label{rule:a} for an $\alpha$-formula $\alpha\in
            \Tlabel{n_{i}}$, $\Tlabel{n_{i+1}} = \{\Arule{\alpha}\} \cup
            \Tlabel{n_{i}}\setminus \{\alpha\}$;
            
            \item\label{rule:b} for a $\beta$-formula $\beta \in
            \Tlabel{n_{i}}$, $\Tlabel{n_{i+1}} = \{\Brulel{\beta}\} \cup
            \Tlabel{n_{i}} \setminus \{\beta\}$ and there exists another
            branch in $B$ of the form $b' = n_0,
            \dots,n_i,n'_{i+1},\dots,n'_k$ such that $\Tlabel{n'_{i+1}} =
            \{\Bruler{\beta}\} \cup \Tlabel{n_{i}} \setminus \{\beta\}$ ;
            
            \item\label{rule:c} for an existential quantified formula
            $\Lhid{x}{\phi'}
            \in \Tlabel{n_{i}}$, $\Tlabel{n_{i+1}} = \{ \phi'' \} \cup
            \Tlabel{n_{i}} \setminus \{\Lhid{x}{\phi'}\}$ where $\phi''
            \dfn \phi' [y/x]$ with $y$ fresh variable;
            
            \item\label{rule:d} in case $\Tlabel{n_{i}}$ is a set formed
            only by elementary formulas, $\Tlabel{n_{i+1}} =
            \nextop{\Tlabel{n_{i}}}$, where $\nextop{\Phi} \dfn
            \set{\phi}{\Lnext{\phi} \in \Phi} \cup
            \set{\Lneg{\phi}}{\Lneg{\Lnext{\phi}} \in \Phi} \cup (\Phi \cap
            \CSys)$.
        \end{enumerate}    
        }%
    \end{enumerate}
\end{definition}
Rules~\ref{rule:a} and~\ref{rule:b} are standard, replacing $\alpha$ and
$\beta$-formulas with one or two formulas according to the matching pattern
of rules in \smartref{tab:alpha_rules}, except for
Rule~\ref{rule:dist_until} that uses the so-called context $\cntx{\Gamma}$,
which is defined in the following.  The $\nextop{}$ operator used in
Rule~\ref{rule:d} is different from the corresponding one of \pltl{} since
it also preserves the constraint formulas.  This is needed for guaranteeing
correctness in the particular setting of \tccp\ where the store is
monotonic.  Finally, Rule~\ref{rule:c} is specific for the $\Lhid{}$ case:
$\Lhid{x}$ is removed after renaming $x$ with a fresh variable\footnote{The
\csltl{} existential quantification does not correspond to the one of FO
logic.  It serves to model local variables, and $\Lhid{x}{\phi}$ can be
seen just as $\phi$ where the information about $x$ is local.  }.
\begin{definition}\label{def:node-inconsistent}
    A node in the tableau is \emph{inconsistent} if it contains a couple of
    formulas $\phi, \Lneg{\phi}$, or the formula $\Lfalse$, or a constraint
    formula $\Lneg{c'}$ such that the merge $c$ of all the (positive)
    constraint formulas $c_1, \dots, c_n$ in the node (\ie\ $c \dfn c_1
    \CSmerge \dots \CSmerge c_n$) is such that $c \CSimp c'$.  A branch is
    \emph{closed} if it contains an inconsistent node.
\end{definition}
The last condition for inconsistence of a node is particular to the \ccp{}
context.

We now describe the algorithm that automatically builds the \csltl{}
tableau for a given set of formulas $\Phi$ (see \cite{CominiTV14-techrep}
for the pseudocode).  The construction consists in selecting at each step a
branch that can be extended by using $\alpha$ or $\beta$ rules or $\Lhid{}$
elimination.  When none of these can be applied, the $\nextop{}$ operator
is used to pass to the next \emph{stage}.  When dealing with eventualities,
to determine the context $\cntx{\Gamma}$ in Rule~\ref{rule:dist_until}, it
is necessary to \emph{distinguish} the eventuality that is being unfolded
in the path.  Given a node $n$ and $\phi\in L(n)$, $\Gamma :=
L(n)\setminus\{\phi\}$.  Then, when Rule~\ref{rule:dist_until} is applied
to a \emph{distinguished} eventuality, we set $\cntx{\Gamma} \dfn
\lubF{}{}_{\gamma \in \Gamma} \Lneg{\gamma}$; otherwise $\cntx{\Gamma} \dfn
\true$.  The use of contexts is the mechanism to detect the loops that
allows one to mark branches containing eventuality formulas as \emph{open}
or to generate inconsistent nodes and mark branches as \emph{closed}.  A
node is marked as \emph{closed} when it is inconsistent while is marked as
\emph{open} when (1) it is the last node of the branch and contains just
constraint formulas or (2) the branch is cyclic and all the eventualities
in the cycle have been already distinguished.

In order to ensure termination of the algorithm, it is necessary to use a
\emph{fair} strategy to distinguish eventualities, in the sense that every
eventuality in an open branch must be distinguished at some point.  This
assumption and the fact that, given a finite set of initial formulas, there
exists only a finite set of possible labels in a systematic tableau, imply
termination of the tableau construction.  Moreover, the constructed tableau
is sound and complete.  Therefore, to check the validity of a formula of
the form $\Limpl{\psi}{\phi}$, with $\phi = \SF(\mgc{p}{x}{})$ and $\psi =
\FDd{ \D }{\SF} (\mgc{p}{x}{})$, we just have to build the tableau for its
negation $\Tname[\Lneg{(\Limpl{\psi}{\phi})}]$ and check if it is closed or
not.  If it is, we have that $\D$ is abstractly correct.  Otherwise, we can
extract from $\Tname[\Lneg{(\Limpl{\psi}{\phi})}]$ an explicit testimony
$\varphi$ of the abstract incorrectness of $\D$.

The construction of $\psi=\FDd{ \D }{\SF} (\mgc{p}{x}{})$ is linear in the
size of $\D$.  The systematic tableau construction of
$\Lneg{(\Limpl{\psi}{\phi})}$ (from what said in \cite{GaintzarainHLNO09})
has worst case $O(2^{O(2^{|\Lneg{(\Limpl{\psi}{\phi})}|})})$.  However, we
believe that such bound for the worst-case asymptotic behavior is quite
meaningless in this context, since it is not very realistic to think that
the formulas of the specification should grow much (big formulas are
difficult to comprehend and in real situations people would hardly try even
to imagine them).  Moreover, note that tableau explosion is due to nesting
of eventualities and in practice really few eventualities are used in
specifications.  Therefore, in real situations, we do not expect that
(extremely) big tableaux will be built.

\section{Related Work}

A Constraint Linear Temporal Logic is defined in \cite{Valencia05} for the
verification of a different timed concurrent language, called \ntcc{},
which shares with \tccp{} the concurrent constraint nature and the
non-monotonic behavior.  The restricted negation fragment of this logic,
where negation is only allowed for state formulas, is shown to be
decidable.  However, no efficient decision procedure is given (apart from
the proof itself).  Moreover, the verification results are given for the
locally-independent fragment of \ntcc{}, which avoids the non-monotonicity
of the original language.  In contrast, in this work, we address the
problem of checking temporal properties for the full \tccp{} language.

Some model-checking techniques have been defined for \tccp\ in the past
\cite{FalaschiV06,AlpuenteFV05,AlpuenteGPV05,FalaschiPV01}.  It is
worth noting that the notions of correctness and completeness in these
works are defined in terms of $\F[]{\P}$, \ie{} in terms of the
concrete semantics, and therefore their check requires a (potentially
infinite) fixpoint computation.  In contrast, the notions of abstractly
incorrect declarations and abstract uncovered elements are defined in
terms of \emph{just one} application of $\FDd{\P}{}$ to $\SF$.
Moreover, since $\FDd{\P}{}$ is defined compositionally, all the checks
are defined on each \progrule\ in isolation.  Hence, our proposal can
be used with partial \programs.  When a property is falsified, model
checking provides a counterexample in terms of an erroneous execution
trace, leaving to the user the problem of locating the source of the
bug.  On the contrary, we identify the faulty process declaration.

In \cite{FalaschiOPV07}, a first approach to the declarative
debugging of a \ccp{} language is presented.  However, it does
not cover the particular extra difficulty of the
non-monotonicity behavior, common to all timed concurrent
constraint languages.  This makes our approach significantly
different.  Moreover, although they propose the use of \ltl{}
for the specification of properties, their formulation, based
on the depth-k concretization function, complicates the task of
having an efficient implementation.

Finally, this proposal clearly relates to the abstract diagnosis framework
for \tccp{} defined for Galois Insertions \cite{CominiTV11absdiag}.  That
work can compete with the precision of model checking, but its main
drawback is the fact that the abstract domain did not allow to specify
temporal properties in a compact way.  In fact, specifications consisted of
sets of \emph{abstract conditional traces}.  Thus, specifications were big
and unnatural to be written.  The use of temporal logic in this proposal
certainly overcomes this problem.

\section{Conclusion and Future Work}

We have defined an abstract semantics for \tccp{} based on the domain of a
linear temporal logic with constraints.  The semantics is correct \wrt\ the
behavior of the language.

By using this abstract semantics, we have defined a method to validate
\csltl\ formulas for \tccp{} \programs.  Since the abstract semantics
cannot be defined by means of a Galois Connection, we cannot use the
abstract diagnosis framework for \tccp\ defined in
\cite{CominiTV11absdiag}, thus we devised (from scratch) a weak version of
the abstract diagnosis framework based only on a concretization function
$\gamma$.  It works by applying $\FDd{\P}{}$ to the abstract specification
and then by checking the validity of the resulting implications (whether
that computation implies the abstract specification).  The computational
cost depends essentially on the cost of that check of the implication.

We have also presented an automatic decision procedure for the \csltl\
logic, thus we can effectively check the validity of that implication.  We
are currently finishing to implement a proof of concept tool, which is
available online at URL \url{http://safe-tools.dsic.upv.es/tadi/}, that
realizes the proposed instance.  Then we would be able to compare with
other tools and assess the ``real life'' goodness of our proposal.

In the future, we also plan to explore other instances of the method based
on logics for which decision procedures or (semi)automatic tools exists.
This proposal can also be immediately adapted to other concurrent
(non-monotonic) languages (like \tcc\ and \ntcc) once a suitable fully
abstract semantics has been developed.

\bibliographystyle{acmtrans}
\bibliography{new-laura,new-alicia,biblio}

\begin{thebibliography}{}

\bibitem[\protect\citeauthoryear{Alpuente, Falaschi, and Villanueva}{Alpuente
  et~al\mbox{.}}{2005}a]{AlpuenteFV05}
{\sc Alpuente, M.}, {\sc Falaschi, M.}, {\sc and} {\sc Villanueva, A.} 2005a.
\newblock {A Symbolic Model Checker for tccp Programs}.
\newblock In {\em First International Workshop on Rapid Integration of Software
  Engineering Techniques (RISE 2004), Revised Selected Papers}. Lecture Notes
  in Computer Science, vol. 3475. Springer-Verlag, 45--56.

\bibitem[\protect\citeauthoryear{Alpuente, Gallardo, Pimentel, and
  Villanueva}{Alpuente et~al\mbox{.}}{2005}b]{AlpuenteGPV05}
{\sc Alpuente, M.}, {\sc Gallardo, M.}, {\sc Pimentel, E.}, {\sc and} {\sc
  Villanueva, A.} 2005b.
\newblock {A Semantic Framework for the Abstract Model Checking of tccp
  Programs}.
\newblock {\em Theoretical Computer Science\/}~{\em 346,\/}~1, 58--95.

\bibitem[\protect\citeauthoryear{Alpuente, Gallardo, Pimentel, and
  Villanueva}{Alpuente et~al\mbox{.}}{2006}]{AlpuenteGPV06}
{\sc Alpuente, M.}, {\sc Gallardo, M.}, {\sc Pimentel, E.}, {\sc and} {\sc
  Villanueva, A.} 2006.
\newblock {Verifying Real-Time Properties of tccp Programs}.
\newblock {\em Journal of Universal Computer Science\/}~{\em 12,\/}~11,
  1551--1573.

\bibitem[\protect\citeauthoryear{Clarke and Emerson}{Clarke and
  Emerson}{1981}]{ClarkeE81}
{\sc Clarke, E.~M.} {\sc and} {\sc Emerson, E.~A.} 1981.
\newblock Design and synthesis of synchronization skeletons using
  branching-time temporal logic.
\newblock In {\em Logic of Programs}, {D.~Kozen}, Ed. Lecture Notes in Computer
  Science, vol. 131. Springer, 52--71.

\bibitem[\protect\citeauthoryear{Comini, Titolo, and Villanueva}{Comini
  et~al\mbox{.}}{2011}]{CominiTV11absdiag}
{\sc Comini, M.}, {\sc Titolo, L.}, {\sc and} {\sc Villanueva, A.} 2011.
\newblock {Abstract Diagnosis for Timed Concurrent Constraint programs}.
\newblock {\em Theory and Practice of Logic Programming\/}~{\em 11,\/}~4-5,
  487--502.

\bibitem[\protect\citeauthoryear{Comini, Titolo, and Villanueva}{Comini
  et~al\mbox{.}}{2013a}]{CominiTV13semTR}
{\sc Comini, M.}, {\sc Titolo, L.}, {\sc and} {\sc Villanueva, A.} 2013a.
\newblock {A Condensed Goal-Independent Bottom-Up Fixpoint Modeling the
  Behavior of \textsf{tccp}}.
\newblock Tech. rep., DSIC, Universitat Polit\`ecnica de Val\`encia. Available
  at \url{http://riunet.upv.es/handle/10251/34328}.

\bibitem[\protect\citeauthoryear{Comini, Titolo, and Villanueva}{Comini
  et~al\mbox{.}}{2013b}]{CominiTV13-WLPE}
{\sc Comini, M.}, {\sc Titolo, L.}, {\sc and} {\sc Villanueva, A.} 2013b.
\newblock {Towards an Effective Decision Procedure for LTL formulas with
  Constraints}.
\newblock In 23rd Workshop on Logic-based methods in Programming Environments
  (WLPE 2013). {\em CoRR\/}~{\em abs/1308.2055}.

\bibitem[\protect\citeauthoryear{Comini, Titolo, and Villanueva}{Comini
  et~al\mbox{.}}{2014}]{CominiTV14-techrep}
{\sc Comini, M.}, {\sc Titolo, L.}, {\sc and} {\sc Villanueva, A.} 2014.
\newblock {Abstract Diagnosis for \tccp\ using a Linear Temporal Logic}.
\newblock Tech. rep., Universitat Polit\`ecnica de Val\`encia. Available at
  \url{http://riunet.upv.es/handle/10251/8351}.

\bibitem[\protect\citeauthoryear{de~Boer, Gabbrielli, and Meo}{de~Boer
  et~al\mbox{.}}{2000}]{deBoerGM99}
{\sc de~Boer, F.~S.}, {\sc Gabbrielli, M.}, {\sc and} {\sc Meo, M.~C.} 2000.
\newblock {A Timed Concurrent Constraint Language}.
\newblock {\em Information and Computation\/}~{\em 161,\/}~1, 45--83.

\bibitem[\protect\citeauthoryear{de~Boer, Gabbrielli, and Meo}{de~Boer
  et~al\mbox{.}}{2001}]{deBoerGM01}
{\sc de~Boer, F.~S.}, {\sc Gabbrielli, M.}, {\sc and} {\sc Meo, M.~C.} 2001.
\newblock {A Temporal Logic for Reasoning about Timed Concurrent Constraint
  Programs}.
\newblock In {\em TIME '01: Proceedings of the Eighth International Symposium
  on Temporal Representation and Reasoning (TIME'01)}. {IEEE} Computer Society,
  Washington, DC, {USA}, 227.

\bibitem[\protect\citeauthoryear{de~Boer, Gabbrielli, and Meo}{de~Boer
  et~al\mbox{.}}{2002}]{deBoerGM02}
{\sc de~Boer, F.~S.}, {\sc Gabbrielli, M.}, {\sc and} {\sc Meo, M.~C.} 2002.
\newblock {Proving correctness of Timed Concurrent Constraint Programs}.
\newblock {\em CoRR\/}~{\em cs.LO/0208042}.

\bibitem[\protect\citeauthoryear{Falaschi, Olarte, Palamidessi, and
  Valencia}{Falaschi et~al\mbox{.}}{2007}]{FalaschiOPV07}
{\sc Falaschi, M.}, {\sc Olarte, C.}, {\sc Palamidessi, C.}, {\sc and} {\sc
  Valencia, F.~D.} 2007.
\newblock {Declarative Diagnosis of Temporal Concurrent Constraint Programs}.
\newblock In {\em Logic Programming, 23rd International Conference, ICLP 2007,
  Proceedings}, {V.~Dahl} {and} {I.~Niemel{\"a}}, Eds. Lecture Notes in
  Computer Science, vol. 4670. Springer-Verlag, 271--285.

\bibitem[\protect\citeauthoryear{Falaschi, Policriti, and Villanueva}{Falaschi
  et~al\mbox{.}}{2001}]{FalaschiPV01}
{\sc Falaschi, M.}, {\sc Policriti, A.}, {\sc and} {\sc Villanueva, A.} 2001.
\newblock {Modeling concurrent systems specified in a temporal concurrent
  constraint language-I}.
\newblock {\em Electronic Notes in Theoretical Computer Science\/}~{\em 48},
  197--210.

\bibitem[\protect\citeauthoryear{Falaschi and Villanueva}{Falaschi and
  Villanueva}{2006}]{FalaschiV06}
{\sc Falaschi, M.} {\sc and} {\sc Villanueva, A.} 2006.
\newblock Automatic verification of timed concurrent constraint programs.
\newblock {\em Theory and Practice of Logic Programming\/}~{\em 6,\/}~3,
  265--300.

\bibitem[\protect\citeauthoryear{Gaintzarain, Hermo, Lucio, and
  Navarro}{Gaintzarain et~al\mbox{.}}{2008}]{GaintzarainHLN08}
{\sc Gaintzarain, J.}, {\sc Hermo, M.}, {\sc Lucio, P.}, {\sc and} {\sc
  Navarro, M.} 2008.
\newblock Systematic semantic tableaux for {PLTL}.
\newblock {\em Electronic Notes in Theoretical Computer Science\/}~{\em 206},
  59--73.

\bibitem[\protect\citeauthoryear{Gaintzarain, Hermo, Lucio, Navarro, and
  Orejas}{Gaintzarain et~al\mbox{.}}{2009}]{GaintzarainHLNO09}
{\sc Gaintzarain, J.}, {\sc Hermo, M.}, {\sc Lucio, P.}, {\sc Navarro, M.},
  {\sc and} {\sc Orejas, F.} 2009.
\newblock {Dual Systems of Tableaux and Sequents for PLTL}.
\newblock {\em The Journal of Logic and Algebraic Programming\/}~{\em 78,\/}~8,
  701--722.

\bibitem[\protect\citeauthoryear{Manna and Pnueli}{Manna and
  Pnueli}{1992}]{MannaP92}
{\sc Manna, Z.} {\sc and} {\sc Pnueli, A.} 1992.
\newblock {\em The temporal logic of reactive and concurrent systems -
  specification}.
\newblock Springer.

\bibitem[\protect\citeauthoryear{Palamidessi and Valencia}{Palamidessi and
  Valencia}{2001}]{PalamidessiV-CP2001}
{\sc Palamidessi, C.} {\sc and} {\sc Valencia, F.~D.} 2001.
\newblock {A Temporal Concurrent Constraint Programming Calculus}.
\newblock In {\em 7th International Conference on Principles and Practice of
  Constraint Programming (CP'01)}. Lecture Notes in Computer Science, vol.
  2239. Springer, 302--316.

\bibitem[\protect\citeauthoryear{Queille and Sifakis}{Queille and
  Sifakis}{1982}]{QueilleS82}
{\sc Queille, J.~P.} {\sc and} {\sc Sifakis, J.} 1982.
\newblock {Specification and verification of concurrent systems in CESAR}.
\newblock In {\em Symposium on Programming}, {M.~Dezani-Ciancaglini} {and}
  {U.~Montanari}, Eds. Lecture Notes in Computer Science, vol. 137. Springer,
  337--351.

\bibitem[\protect\citeauthoryear{Saraswat}{Saraswat}{1989}]{Saraswat89phd}
{\sc Saraswat, V.~A.} 1989.
\newblock {Concurrent Constraint Programming Languages}.
\newblock Ph.D. thesis, Carnegie-Mellon University.

\bibitem[\protect\citeauthoryear{Saraswat}{Saraswat}{1993}]{Saraswat93}
{\sc Saraswat, V.~A.} 1993.
\newblock {\em {Concurrent Constraint Programming}}.
\newblock The MIT Press, Cambridge, Mass.

\bibitem[\protect\citeauthoryear{Valencia}{Valencia}{2005}]{Valencia05}
{\sc Valencia, F.~D.} 2005.
\newblock Decidability of infinite-state timed {CCP} processes and first-order
  {LTL}.
\newblock {\em Theoretical Computer Science\/}~{\em 330,\/}~3, 577--607.

\end{thebibliography}

\end{document}